\documentclass[%
reprint,
superscriptaddress,
amsmath,amssymb,amsthm,mathrsfs,
aps,
prl,
]{revtex4-1}

\usepackage{graphicx}
\usepackage{dcolumn}
\usepackage{bm}
\usepackage[usenames]{color}
\usepackage[ 
margin=0.75in,
]{geometry}
\usepackage[normalem]{ulem}

\def\SNFA{Sr$_{1-x}$Na$_{x}$Fe$_2$As$_2$}

\def\Ts{$T_s$}

\def\Rw{$R_w$}

\def\rmid{$r_{\mathrm{mid}}$}
\def\dlt{$\delta$}

\begin{document}

\title{
Quantitative characterization of short-range orthorhombic fluctuations in FeSe through pair distribution function analysis
}

\author{Benjamin A. Frandsen}
\email{benfrandsen@byu.edu}
	\affiliation{ %
	Department of Physics and Astronomy, Brigham Young University, Provo, Utah 84602, USA.
} %
\affiliation{%
	Materials Sciences Division, Lawrence Berkeley National Laboratory, Berkeley, California 94720, USA.
}%

\author{Qisi Wang}
\affiliation{ %
	State Key Laboratory of Surface Physics and Department of Physics, Fudan University, Shanghai 200433, China.
} %

\author{Shan Wu}
\affiliation{ %
	Department of Physics, University of California, Berkeley, California 94720, USA.
} %

\author{Jun Zhao}
\affiliation{ %
	State Key Laboratory of Surface Physics and Department of Physics, Fudan University, Shanghai 200433, China.
} %
\affiliation{ %
	Collaborative Innovation Center of Advanced Microstructures, Nanjing 210093, China.
} %

\author{Robert J. Birgeneau}
\affiliation{ %
	Department of Physics, University of California, Berkeley, California 94720, USA.
} %
\affiliation{%
	Materials Sciences Division, Lawrence Berkeley National Laboratory, Berkeley, California 94720, USA.
}%
\affiliation{ %
	Department of Materials Science and Engineering, University of California, Berkeley, California 94720, USA.
} %

\begin{abstract}
Neutron and x-ray total scattering measurements have been performed on powder samples of the iron chalcogenide superconductor FeSe. Using pair distribution function (PDF) analysis of the total scattering data to investigate short-range atomic correlations, we establish the existence of an instantaneous, local orthorhombic structural distortion attributable to nematic fluctuations that persists well into the high-temperature tetragonal phase, at least up to 300~K and likely to significantly higher temperatures. This short-range orthorhombic distortion is correlated over a length scale of about 1~nm at 300~K and grows to several nm as the temperature is lowered toward the long-range structural transition temperature. In the low-temperature nematic state, the local instantaneous structure exhibits an enhanced orthorhombic distortion relative to the average structure with a typical relaxation length of 3~nm. The quantitative characterization of these orthorhombic fluctuations sheds light on nematicity in this canonical iron-based superconductor.
\end{abstract}

\maketitle

Understanding the role of electronic nematicity in iron-based superconductors (FeSCs) remains an outstanding goal among researchers in the field. Characterized by an electronically-driven splitting of the $d_{xz}$ and $d_{yz}$ bands that reduces the original $C_4$ rotational symmetry to $C_2$, nematic order manifests as a tetragonal-to-orthorhombic structural phase transition and a significant anisotropy in the spin susceptibility, which almost always triggers long-range magnetic order~\cite{pagli;np10,ferna;np14,hoson;pc15,si;nrm16}. Because optimal superconductivity occurs near a nematic instability in diverse families of FeSCs, nematicity and superconductivity are thought to share a profound connection~\cite{shiba;arocmp14,chen;nsr14,leder;prl15,bohme;crp16,kuo;s16,chubu;prx16,matsu;nc17,ferna;rpp17,leder;pnas17}. Nevertheless, a comprehensive understanding of the origin of electronic nematicity and its precise relationship to superconductivity remains elusive, in part due to the complexity introduced by the coupling between nematicity, orbital order, and magnetic order.

\begin{figure}
	\includegraphics[width=65mm]{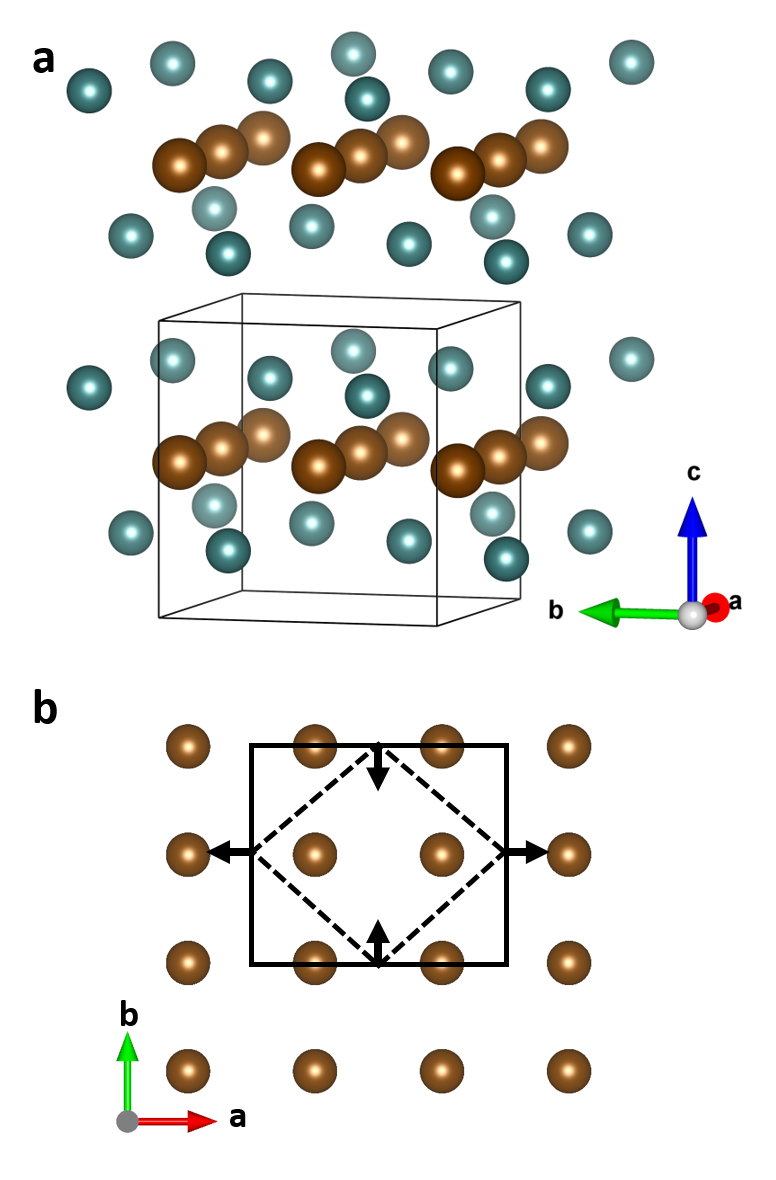}
	\caption{\label{fig:struc} (a) Crystal structure of FeSe, with Fe atoms shown in brown and Se atoms in turquoise. The solid lines show the unit cell of the orthorhombic structure (space group $Cmma$). (b) View of the distorted iron sublattice within each layer of the crystal structure. The solid rectangle represents the projection of the orthorhombic unit cell onto the Fe plane, while the dashed rhombus shows the distorted tetragonal unit cell. The black arrows indicate the strain directions leading to the distortion. The magnitude of the distortion has been exaggerated for clarity. The red and green arrows indicate the directions of $a$ and $b$ orthorhombic lattice vectors, respectively.}
	
\end{figure}

Iron selenium (FeSe) is unique among FeSCs in that it possesses the simplest crystal structure and does not exhibit long-range magnetic order in the low-temperature nematic phase~\cite{marga;cc08,pomja;prb09}. As such, it is a promising system in which to investigate electronic nematicity. Indeed, FeSe has been the object of extensive research efforts, focused primarily on the orbital and magnetic degrees of freedom and their interrelationships with nematicity and superconductivity~\cite{nakay;prl14,shimo;prb14,watso;prb15,chubu;prb15,wang;nm16,wang;np15,baek;nm15,yamak;prx16,wang;nc16,yi;arxiv19}. The high-temperature crystal structure is described by the tetragonal space group $P4/nmm$ and transforms to orthorhombic $Cmma$ in the nematic state below $T_s \approx 90$~K, resulting in the square lattice of Fe atoms present at high temperature deforming into a rectangular lattice at low temperature. The crystal structure of FeSe is displayed in Fig.~\ref{fig:struc}(a), with the distorted Fe sublattice shown schematically in panel (b). In terms of the orthorhombic unit cell, the orthorhombic distortion corresponds to a nonzero difference between the length of the $a$ and $b$ lattice vectors, which can be quantified by the dimensionless orthorhombicity parameter $\delta = (a-b)/(a+b)$. For FeSe, the maximum distortion of the average structure at low temperature corresponds to $\delta \approx 0.002$~\cite{pomja;prb09}.

In FeSe and other systems, valuable insight can be gained through studies of not only the static nematic phase, but also fluctuations and other signatures of the nematic phase at temperatures above \Ts~\cite{chu;s10,chu;s12,rosen;np14,bohme;prl14,iye;jpsj15,bohme;crp16,kuo;s16,galla;prl16,hosoi;pnas16,massa;pnas16,kretz;np16,liu;prl16,xu;prb16,luo;qm17,palms;prb17,wang;prb17,baek;nc18,wang;nc18}. These fluctuations may contain clues about the origin of nematicity and its connection to superconductivity. The relatively large characteristic energy associated with the nematicity ($\sim$50~meV~\cite{yi;arxiv19}) indicates that the nematic order parameter drives the structural phase transition, accomplished through linear coupling to the orthorhombic order parameter  \cite{chu;s12,ferna;np14,bohme;prl15}. Fluctuating nematicity therefore results in fluctuating orthorhombic distortions of the lattice. Consequently, studying fluctuations of the orthorhombic order parameter can provide insights into fluctuations of the nematic order parameter.

Short-range, fluctuating orthorhombicity has recently been observed in the hole-doped pnictide system \SNFA\ through pair distribution function (PDF) analysis of neutron and x-ray total scattering data~\cite{frand;prl17,frand;prb18}. By Fourier transforming the total scattering pattern, which includes both Bragg scattering from long-range structural correlations and diffuse scattering from short-range correlations, the PDF method enables quantitative refinement of the instantaneous atomic structure on short length scales in real space~\cite{egami;b;utbp12}. The previous work on \SNFA\ revealed the presence of short-range orthorhombic distortions correlated over a 2-3~nm length scale that persist up to remarkably high temperatures ($\sim$500~K for the parent compound) and survive even in heavily doped compounds exhibiting optimal superconductivity. However, the few published PDF studies of 11-type iron chalcogenide systems~\cite{hu;prb09,louca;prb10} have not examined nematic fluctuations, leaving a gap in the experimental literature.

We address that need here by presenting a thorough PDF analysis of x-ray and neutron total scattering measurements of FeSe. We observe short-range, nanometer-scale orthorhombic distortions of the instantaneous structure at least up to room temperature, which grow in correlation length upon cooling and persist in the static nematic phase in the form of enhanced local orthorhombicity relative to the long-range distortion of the average structure. In addition to providing real-space confirmation and a quantitative characterization of enhanced orthorhombic fluctuations in FeSe, these results further establish the relevance of fluctuating nematicity across multiple families of FeSCs.

Single crystals of FeSe were grown under a permanent temperature gradient ($\sim$400~$^{\circ}$C to 330~$^{\circ}$C) using KCl-AlCl$_3$ flux~\cite{wang;nc16}. Samples with a total mass of 2~g were gently ground with a mortar and pestle inside an argon-filled glove box for 10 minutes to produce a powder used for neutron total scattering measurements. A second, identically prepared powder of mass 0.5~g, was produced for equivalent x-ray characterization.

The neutron total scattering measurements were performed on the General Materials Diffractometer (GEM), a time-of-flight instrument at the ISIS Neutron Source of Rutherford Appleton Laboratory~\cite{willi;pb97}. The sample was placed in a closed cycle refrigerator to control the temperature. Scattering data were collected for 12 hours for each temperature point measured. The raw total scattering structure function $S(Q)$ was reduced and transformed to the real-space PDF using the Gudrun package~\cite{soper;jac11} installed on the beamline computers. A maximum momentum transfer of 35.0~\AA$^{-1}$ was used for the Fourier transform, corresponding to a real-space resolution of $\pi/Q_{max} = $ 0.0898~\AA. Since the GEM instrument does not utilize energy analysis of the scattered beam, the experimental PDF data contain information about structural correlations on time scales as short as 10$^{-13}$~s.

The x-ray total scattering measurements were conducted on the Pair Distribution Function beamline (28-ID-1) at the National Synchrotron Light Source II located at Brookhaven National Laboratory. The incident beam had a wavelength of 0.1867~\AA. Background and calibration measurements were performed according to standard protocols, as was also the case for the neutron PDF data. A Perkin-Elmer area detector was used to collect the raw diffraction data, which were then integrated, reduced, and transformed using the Fit2D~\cite{hamme;hpr96} and xPDFsuite~\cite{yang;arxiv15} software packages. The maximum momentum transfer used for the Fourier transform was 25.0~\AA$^{-1}$, leading to PDF data with somewhat lower real-space resolution (0.126~\AA) compared to the neutron PDF data. Additionally, the low $Q$-space resolution at the x-ray beamline imposes a damping envelope on the real-space PDF data, restricting reliable data to below $\sim$50~\AA. Due to the high intensity of the x-ray beam, a collection time of 3 minutes per temperature point was sufficient, enabling a thorough study of the temperature dependence. Given the large energy scale of the incident x-rays compared to electronic energy scales, the x-ray PDF data reflect the true instantaneous atomic structure of the sample. The neutron and x-ray PDF data were analyzed and modeled using the DiffPy-CMI package~\cite{juhas;aca15}. All refinements were carried out on the Nyquist sampling grid.

\begin{figure}
	\includegraphics[width=85mm]{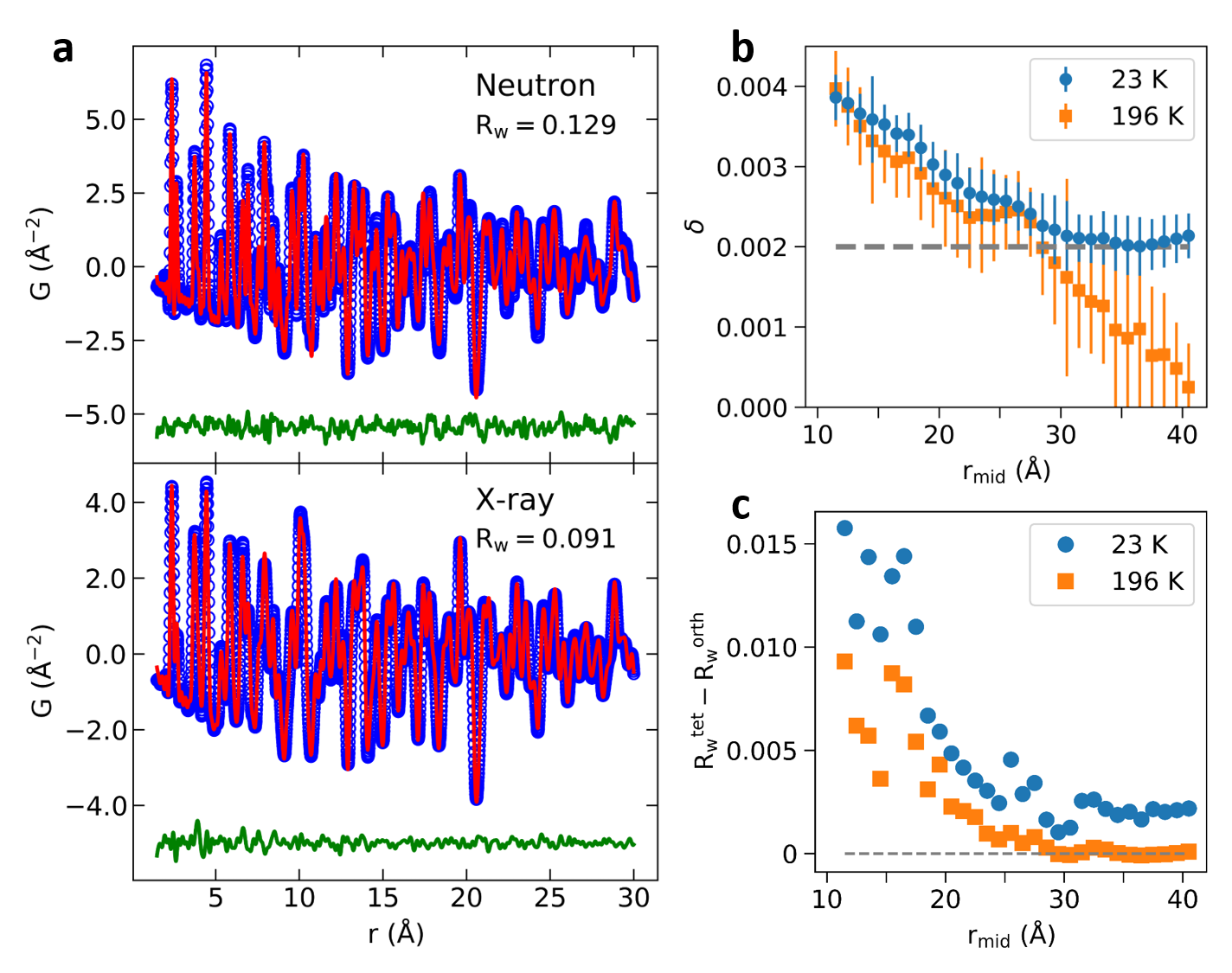}
	\caption{\label{fig:neutronResults} (a) Fit to the neutron (top) and x-ray (bottom) PDF data using the orthorhombic structural model. The neutron and x-ray data were collected at 23~K and 6~K, respectively. The blue circles represent the data, the red curve the calculated PDF from the refined structural model, and the green curve the fit residual, offset vertically for clarity. (b) Refined orthorhombicity \dlt\ determined as a function of the midpoint of the fitting range for the data collected at 196~K (orange squares) and 23~K (blue circles). The steady decrease of \dlt\ to zero for the 196~K data reflects the short-range nature of the orthorhombic distortion at that temperature, while the plateau around 0.002 at high \rmid\ for the 23~K data indicates a long-range structural distortion. The horizontal dashed line indicates the magnitude of the low-temperature orthorhombic distortion in the average structure as determined by conventional Rietveld refinement. Error bars show the estimated standard deviations of the refined parameters. (c) Difference in $R_w$ between the tetragonal and orthorhombic models as a function of fitting range for the data collected at 23~K (blue circles) and 196~K (orange squares).}
	
\end{figure}

We begin with the neutron PDF analysis. Due to the time required to obtain high-quality data, we focused on just two temperatures, 196~K (in the high-temperature tetragonal phase) and 23~K (deep in the nematic phase). A fit to the data at 23~K using the $Cmma$ orthorhombic structural model and the nominal 1:1 composition is shown in Fig.~\ref{fig:neutronResults}(a), along with an x-ray PDF fit to be discussed subsequently. 
The refined orthorhombicity is $\delta$~=~0.0023(2) as determined by a fit to the neutron data over the range 1.5 -- 50~\AA\ , consistent with the expected value of $\sim$0.002 based on conventional Rietveld refinement. Refining the tetragonal model yields a poorer fit ($R_w = 0.136$ compared to 0.129 for the orthorhombic model), confirming that the data are sufficient to resolve the orthorhombicity reliably. The estimated uncertainty of the $R_w$ values is below 0.001 as determined by bootstrapping calculations, indicating that the difference between the two models is significant. Further details are provided in the Supplementary Information. 

An equivalent refinement of the orthorhombic structural model against the data collected at 196~K yields a best-fit $\delta$ that is very close to zero for a fitting range of 1.5 -- 50~\AA. This is not surprising, given that the long-range crystallographic structure of FeSe is tetragonal at this temperature. However, refining the same structural model over a shorter fitting range from 1.5 -- 21.5~\AA\ results in a large $\delta$ of $\sim$0.004. Additionally, refining the purely tetragonal model over the same range yields a significantly worse fit, with $R_w = 0.128$ compared to 0.119 for the orthorhombic model. This indicates that the instantaneous local structure up to 21.5~\AA\ is orthorhombically distorted, but when averaged over longer distances up to 50~\AA, the distortion is greatly reduced. This local orthorhombicity is similar to that established by previous PDF measurements of \SNFA~\cite{frand;prl17,frand;prb18}. We attribute this local orthorhombic distortion to short-range nematicity in the high-temperature phase of FeSe.

To examine the length scale of the short-range orthorhombicity at 196~K, we performed a series of refinements of the orthorhombic model over a sliding data window ranging from [1.5~\AA\ - 21.5~\AA] to [30.5~\AA\ - 50.5~\AA] in steps of 1~\AA. The refined structural model for each fitting window represents the best-fit structure on that length scale. A width of 20~\AA\ for the fitting ranges was chosen because this provides enough data to ensure robust convergence of the fits but is not such a wide range that it averages away any local structure. To reduce the correlation between fitting parameters, we enforced tetragonal symmetry of the atomic displacement parameters (ADPs) within the orthorhombic model, so that the only additional free parameter in the orthorhombic model relative to the tetragonal model was the orthorhombicity $\delta$. We refined $\delta$ directly by parameterizing the in-plane lattice constants for the orthorhombic model as $a = a_{\mathrm{mid}}(1+\delta)$ and $b = a_{\mathrm{mid}}(1-\delta)$, where $a_{\mathrm{mid}} = (a+b)/2$ is equivalent to the tetragonal in-plane lattice constant. In addition, we refined the purely tetragonal model over the same fitting ranges to allow a comparison of the goodness of fit for each model.

The results of this procedure are displayed in Fig.~\ref{fig:neutronResults}(b) and (c). In panel (b), the orange squares represent the refined values of $\delta$ at 196~K as a function of \rmid, the midpoint of the fitting range. At low \rmid, $\delta$ is quite large, exceeding even 0.002 corresponding to the long-range orthorhombic distortion at low temperature (shown as the dashed horizontal line). However, $\delta$ steadily decreases as the fitting range increases, becoming statistically indistinguishable from zero for \rmid\ between 35 and 40~\AA. Similarly, the orange squares in panel (c) show the difference in $R_w$ obtained from the tetragonal and orthorhombic models; the higher this difference, the greater the improvement of the orthorhombic model over the tetragonal model. The difference in \Rw\ approaches 0.01 for the low-$r$ fits but steadily decreases as the fitting range increases, eventually reaching zero around 30~\AA, close to where the refined orthorhombicity reaches zero within error bars. These results provide a direct illustration of the short-range nature of the orthorhombic distortion at 196~K and reveal a length scale of approximately 3~nm. We note that the characteristic length scale revealed by PDF analysis may differ somewhat from the true correlation length of the nematic order parameter. 

The blue symbols in Fig.~\ref{fig:neutronResults}(b) show the corresponding results for \dlt\ obtained from fits performed on the data collected at 23~K. Interestingly, the refined values of \dlt\ are nearly identical for both temperatures at low $r$, where we once again observe an enhanced local orthorhombicity that decreases as the fitting range is increased. Beyond about \rmid~=~30~\AA, the low-temperature value of \dlt\ plateaus at the expected value corresponding to the long-range distortion of the average structure. Likewise, the blue circles in Fig.~\ref{fig:neutronResults}(c) mark the difference in \Rw\ between the tetragonal and orthorhombic models as a function of fitting range, showing a steady decrease from about 0.015 at low $r$ to a plateau around 0.003 at high $r$. The enhanced short-range orthorhombicity may suggest a scenario of incomplete nematic ordering with significant nematic fluctuations persisting even at 23~K.

The neutron PDF results presented so far establish the existence of a short-range, instantaneous orthorhombic distortion correlated over approximately 3~nm that exists well into the high-temperature tetragonal phase, as well as locally enhanced orthorhombicity on a similar length scale deep in the nematically ordered state. To obtain a more complete temperature dependence of the local structure, we turn to the x-ray PDF data, which were collected between 6~K and 300~K in steps of 6~K. A representative x-ray PDF fit is shown in Fig.~\ref{fig:neutronResults}(a), for which the orthorhombic model (shown) yields $R_w$ = 0.091 compared to 0.096 for the tetragonal model (not shown). For the PDF data collected at each temperature, we performed the same type of sliding-range fits done for the neutron data. The output of this procedure was a set of best-fit orthorhombic and tetragonal structural models corresponding to a dense (\rmid, $T$) grid.

\begin{figure}
	\includegraphics[width=70mm]{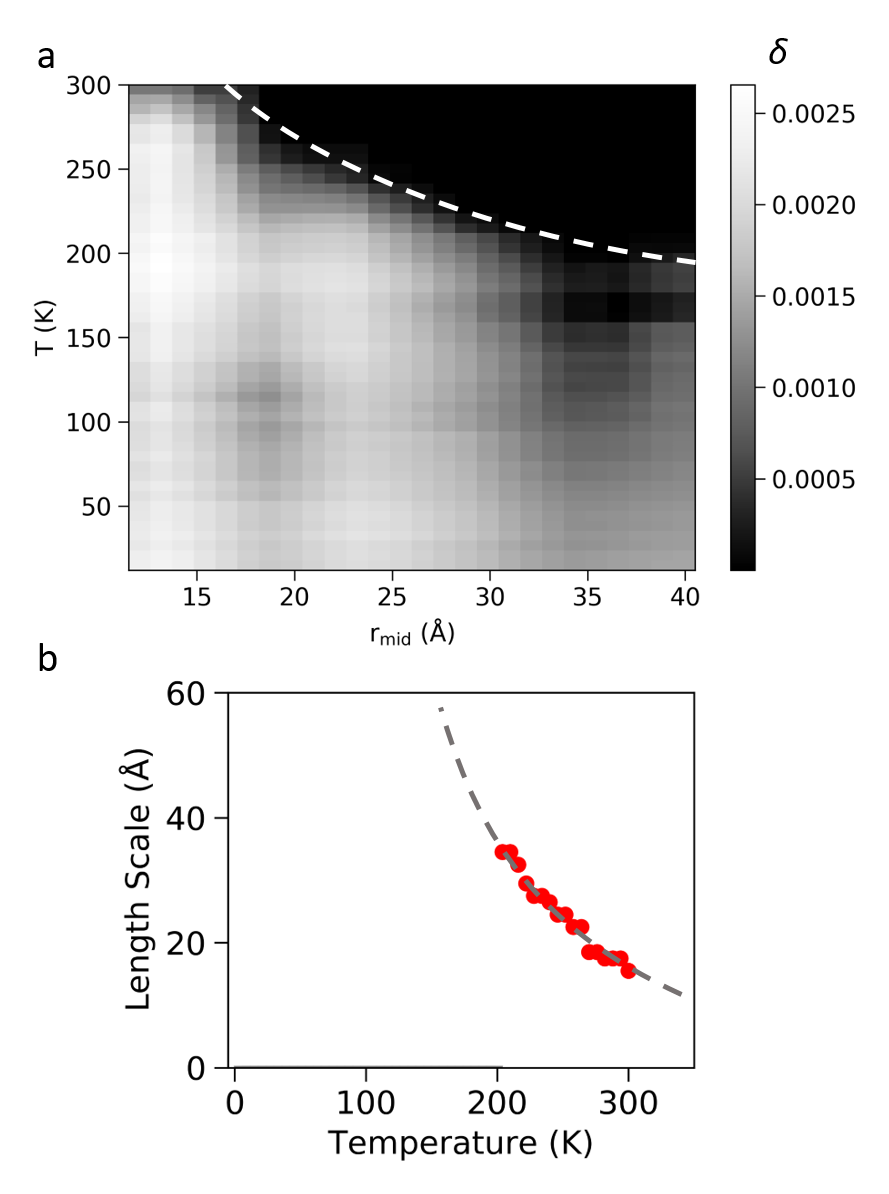}
	\caption{\label{fig:xray} (a) Grayscale plot of the orthorhombicity \dlt\ determined from x-ray PDF analysis of FeSe. The horizontal axis represents the midpoint of the fitting window, the vertical axis temperature. The white dashed curve represents an estimate of the length scale of nematic correlations. (b) Characteristic length scale of the local orthorhombic distortion in FeSe as a function of temperature. Below 200~K, the length scale exceeds the sensitivity of the PDF data. The gray dashed curve is a guide to the eye, corresponding to the white dashed curve in (a).}
	
\end{figure}

In Fig.~\ref{fig:xray}(a), we present a grayscale plot showing the refined value of \dlt\ as a function of \rmid\ and $T$, with the brightness corresponding to the value of \dlt\ as quantified by the color bar.
This plot provides a visual representation of the evolution of the local orthorhombic distortion with temperature. The distortion is nonzero even at 300~K, but extends only up to fitting ranges with \rmid\ between 15 and 20~\AA. Over longer real-space distances, the refined orthorhombicity is zero. As the temperature is lowered, \dlt\ remains nonzero to increasingly high \rmid. Indeed, below 200~K, even the longest fitting range of 30.5 -- 50.5~\AA\ yields nonzero \dlt. We were unable to conduct reliable fits over longer $r$ ranges due to the significant damping of the PDF data beyond $\sim$50~\AA\ originating from limited $Q$ resolution in the raw diffraction data. The average uncertainty of the refined value of \dlt\ is 5.2$\times 10^{-4}$.

We illustrate the growth of the orthorhombic distortion in Fig.~\ref{fig:xray}(b), which displays the temperature dependence of the estimated orthorhombic length scale (defined here as the value of \rmid\ of the fitting range for which \dlt\ first reaches zero within the error bars). We see a steady increase as the temperature is lowered until 200~K, below which the characteristic length scale exceeds the feasible range of the experimental data, so no additional points have been plotted. The dashed gray curve is a guide to the eye illustrating the trend with temperature. This is also represented as the white dashed curve in Fig.~\ref{fig:xray}(a). 

As the temperature is lowered below \Ts~=~90~K, \dlt\ increases in magnitude in the high-$r$ range, while the low-$r$ range remains relatively unchanged. As with the neutron PDF data, the low-$r$ orthorhombicity is enhanced relative to that at high-$r$, with a similar relaxation length of $\sim$30~\AA. We note that the values of \dlt\ obtained from the x-ray PDF data are systematically slightly lower than those obtained from the neutron data, which is most likely a consequence of the lower real-space resolution of the x-ray data and/or a difference in time scale sensitivity between the two measurements resulting from differing effective energy integration windows. The $r$- and $T$-dependent behavior of the refined \dlt\ values is confirmed by the refined ADP values, as discussed in the Supplementary Information.

Taken together, the results of the x-ray PDF analysis confirm and extend the neutron PDF results, establishing the existence of a nanometer-scale orthorhombic distortion of the instantaneous local atomic structure that survives well into the high-temperature tetragonal phase. This local distortion, a manifestation of nematic fluctuations, grows in correlation length as the temperature is lowered toward the long-range nematic ordering temperature \Ts. Moreover, it persists below \Ts\ in the form of an enhanced orthorhombic distortion at low $r$ compared to high $r$, with an approximate relaxation length of 30~\AA\ until it reaches the magnitude of the long-range distortion.

The PDF analysis reported here represents direct evidence of short-range, instantaneous orthorhombic distortions of the FeSe lattice, and further provides quantitative insights into the amplitude, length scale, and temperature dependence of these fluctuations. The robustness of these fluctuations at elevated temperatures is especially notable. Although we conducted no measurements above 300~K, extrapolation of the length-scale trend in Fig.~\ref{fig:xray}(b) suggests that these fluctuations may persist up to temperatures as high as 500 -- 600~K. Interestingly, this is qualitatively consistent with the 50-meV (580~K) splitting of the $d_{xz}$ and $d_{yz}$ bands at low temperature revealed by photoemission spectroscopy, which reflects the underlying nematic energy scale in FeSe~\cite{nakay;prl14,shimo;prb14,yi;arxiv19}. In this sense, the presence of local nematic symmetry breaking at room temperature and above may be a natural consequence of the large nematic energy scale in FeSe.

The present results are quite similar to those from an earlier PDF analysis of \SNFA, which established the presence of short-range orthorhombic distortions on a length scale of 2-3~nm up to temperatures as high as 500~K in undoped SrFe$_2$As$_2$~\cite{frand;prl17,frand;prb18}.  The similarities confirm the relevance of robust local orthorhombicity attributable to nematic fluctuations across multiple families of FeSCs. Further systematic PDF characterization of short-range orthorhombicity in other iron pnictide and chalcogenide systems will likely prove to be illuminating in the ongoing effort to establish the origin of nematicity and its connection to superconductivity. In particular, comparing the characteristic length scale and temperature dependence of the short-range orthorhombicity, or equivalently, nematicity revealed by PDF with the corresponding attributes of other fluctuating electronic degrees of freedom probed by complementary techniques, such as inelastic neutron scattering, will be highly informative. Earlier inelastic neutron scattering studies of the antiferromagnetic spin fluctuations in FeSe already provide significant insight into the interplay between spin and nematic fluctuations~\cite{wang;nm16,wang;nc16}, motivating future work focused on the real-space extent and temperature dependence of the fluctuating antiferromagnetism.

We note that during the preparation of this manuscript, we became aware of a similar work involving x-ray PDF analysis of FeSe that appeared on the arXiv preprint server~\cite{koch;prb19}. The conclusions between the two works are largely consistent.

\textbf{Acknowledgements}

We acknowledge Dung-Hai Lee, Ming Yi, and Yu Song for valuable discussions. We thank Dave Keen for his assistance with the neutron PDF experiments at ISIS and Milinda Abeykoon and Eric Dooryhee for their help with the x-ray PDF work at NSLS-II. Work at Lawrence Berkeley National Laboratory was funded by the U.S. Department of Energy, Office of Science, Office of Basic Energy Sciences, Materials Sciences and Engineering Division under Contract No. DE-AC02-05-CH11231 within the Quantum Materials Program (KC2202). Q. W. and J. Z. were supported by the Innovation Program of Shanghai Municipal Education Commission (grant number 2017-01-07-00-07-E00018), the National Natural Science Foundation of China (grant number 11874119), the Ministry of Science and Technology of China (Program 973: 2015CB921302), and the National Key R\&D Program of the MOST of China (grant number 2016YFA0300203). Use of the National Synchrotron Light Source II at Brookhaven National Laboratory, was supported by DOE-BES under Contract No. DE-SC0012704.

\end{document}